\documentclass[11pt,psfig]{article}
\usepackage[latin1]{inputenc}
\usepackage{amsmath}
\usepackage{amsfonts}
\usepackage{amssymb}
\usepackage{geometry}
\usepackage{algorithmic}
\usepackage{algorithm}
\usepackage{tikz} %%Generate vector graphics from within LaTeX
\usepackage{graphicx}
\usepackage{pstricks,pst-grad}
\usepackage{hyperref}
\usepackage{pifont}
\usepackage{cite}
\usepackage{amsthm}
\usepackage{amsmath}
\usepackage{amsfonts}
\usepackage{epsfig}
\usepackage{balance}

\newtheorem{thm}{Theorem}

\newtheorem{prop}{Proposition}
\newtheorem{corollary}{Corollary}
\newtheorem{lem}{Lemma}
\newtheorem{rem}{Remark}

\newcommand{\LL}{\hat{L}}
\newcommand{\UU}{\hat{U}}
\newcommand{\ZZ}{\bar{Z}}

\begin{document}
\begin{center}

{\Huge \textbf{One bank problem in the federal funds market}}

%\author{Cristina Canepa and Traian A. Pirvu}
\mbox{}\\[0pt]
\vspace{0.2cm} Cristina Canepa\\[0pt]  
Department of Mathematics \& Statistics\\[0pt]
McMaster University \\[0pt]
1280 Main Street West \\[0pt]
Hamilton, ON, L8S 4K1\\[0pt]
tpirvu@math.mcmaster.ca \vspace{1cm}

\vspace{0.2cm} Traian A.~Pirvu\\[0pt]  
Department of Mathematics \& Statistics\\[0pt]
McMaster University \\[0pt]
1280 Main Street West \\[0pt]
Hamilton, ON, L8S 4K1\\[0pt]
tpirvu@math.mcmaster.ca \vspace{1cm}
\mbox{}\\[0pt]

\end{center}

%\maketitle

\begin{abstract}
The model in this paper gives a convenient strategy that one bank in the federal funds market can use in order to maximize its profit in a contemporaneous reserve requirement (CRR) regime. The reserve requirements are determined by the demand deposit process, modelled as a Brownian motion with drift. We propose a new model in which the cumulative funds purchases and sales are discounted at possible different rates. We formulate and solve the bank problem of finding the optimal strategy. The model can be extended to involve the bank's asset size and we obtain that, under some conditions, the optimal upper barrier for selling is a linear function of the asset size. As a consequence, the net purchase amount turns to be linear in the asset size.
\end{abstract}
\section{Introduction}

One of the threats perceived as being brought by a financial crisis on economy is the lack of liquidity. Because of this threat, in 2008, big banks received help from the Federal Reserve Bank through public money, whereas small banks were let to default or to be acquired by other banks. It was argued that the bailouts were necessary because a big bank's default might be followed by a lack of liquidity in the market and, thus, by a cascade of defaults. However, letting small banks to fail and helping the big ones to become bigger leads to a market which is even more concentrated and to bigger liquidity problems in a future financial crisis, as the Nobel prize winner J. Stieglitz points out in [11]. It is of great interest, therefore, a thorough study of the connection between bank's market concentration and some measure of liquidity. Our model puts in evidence the net purchase amount as a function of the bank's assets.

Since the link between liquidity and market concentration is important both for the regulator and for the market players, we take into account macro policies regarding bank's activities. The latter are first implemented in the federal funds market, which incorporates all the transactions involving federal funds. Therefore, it is worthwhile to take in consideration the particularities of the federal funds market, even in a stylized way. An interesting problem would be if the model could put in evidence this so-called small bank- large bank dichotomy.

The one bank problem from this paper involves an economy with only one bank and the Federal Reserve Bank. We assume that the bank is a price-taker in the federal funds market and can obtain sufficient funds from the Federal Reserve Bank. The bank's task is to derive an optimal transaction amount to minimize the cost of buying and selling (actually borrowing and lending) overnight federal funds, while meeting the reserve requirements in a CRR regime. We ignore long-term time deposits and we assume that the bank's only source of funds are demand deposits and federal funds. A percent of the deposits need to be kept as reserves in the bank's account with the Fed. The difference between the deposits and the required reserves represent the excess reserves, which can be used by the bank for the transactions in the federal funds market. We consider that during the business day, the bank can increase/decrease its level of federal funds through direct transactions, which involve transaction costs.

The model of this paper extends the optimal control problem of [2] and [6], by allowing for the transactions to be discounted at possibly different rates. We formulate and solve the problem by giving the optimal value function and the optimal strategy. A different approach is to use the entropy maximization as in [9], [10].

As in [1] and [2], our model gives an optimal net purchase amount as an output, using the martingale/ supermartingale principle (see e.g. [8]) and the double Skorokhod formula. The model is based on the asset size as in [3]. We obtain that the net purchase amount is increasing in the asset size. Anecdotally, large banks were known to be buyers of funds whereas small, risk averse banks were known to be sellers of funds (see [4], [12]). Therefore, our model puts in evidence this so called small bank large bank dichotomy.

The paper is structured as follows. In Section 2 we present the model and the main assumptions. In Section 3 we give the problem formulation and present the objective of the paper.
In Section 4 we present the main results and discuss a particular interesting case. The paper ends with an an appendix containing the proofs. 
 
  \section{The model}\label{Descr}
Let us consider the problem of a bank which has an exogenously given demand deposit (net of withdrawals) 
and continuously sells and buys funds, thus lowering or increasing the excess reserves, defined as the 
difference between deposits and required reserves.
   
   The bank is characterized by:
   \begin{enumerate} 
   \item A demand deposit process $(D_{t})_{t \geq 0}$.
   
   \item  A required reserve process $(R_{t})_{t \geq 0}$, where $R_{t}=qD_{t}$.
   
   \item An excess reserve process $X_{t}=(1-q){D}_{t}$.
 \end{enumerate} 

Therefore, modelling the deposits $D$ is equivalent to modelling the excess reserves $X$.

   Let $(\Omega, \textit{F},P_{x})$ be a probability space rich enough to accommodate
a standard, one-dimensional, Brownian motion $B = (B_{t}, 0 \leq t \leq \infty).$ 

Let us consider the problem of a bank which has an exogenous demand deposits (net of withdrawals)
 and continuously sell and buy funds, thus lowering or increasing the demand deposits.
The demand deposits $X = (X_{t}, 0 \leq t \leq \infty)$ are assumed to fluctuate over time as follows \begin{equation}\label{X}
dX_{t}= \mu dt + \sigma dB_{t}.
\end{equation}

 We consider $\textbf{F}=(\textit{F}_{t})_{t\geq0}$ to be the completion of the augmented filtration generated by $X$ (so that $(\textit{F}_{t})$ satisfies the usual conditions).
    
     Therefore, the bank observes nothing except the sample path of $X.$

 \subsection{Policies}

\newtheorem{mydef}{Definition}[section]
 
\begin{mydef} \label{def:policy}
 A policy is defined as a pair of processes  $L$ and $U$ such that
\begin{equation}
\label{adaptLU}
 L, U \qquad \text{are}\qquad\textbf{F}-\text{adapted, right-continuous, increasing and positive.}
\end{equation}
In the context of the federal funds market, $L_{t}$ and $U_{t}$ are the cumulative funds purchases and funds sales (from the central bank) that the bank undertakes up to time $t$, in order to satisfy the reserve requirements and to maximize its profit. Let us take $\lambda_1$ and $\lambda_2,$ $\lambda_1\geq \lambda_2$ be interest rates at which the bank lends and borrows funds. A controlled process associated to the policy ($L,U$) is a process $Z = X+L-U$.
Using formula (\ref{X}) for $X$, we obtain the decomposition of $Z$ into its continuous part and its finite variation part:
\begin{equation}\label{Z}
dZ_{t}= \mu dt + \sigma dB_{t} + dL_{t} - dU_{t}.
\end{equation} 
In our model $Z_{t}$ is the amount of excess funds in the bank's reserve account at time $t.$
The policy ($L,U$) is said to be feasible if
\begin{equation}
L_{0-}=U_{0-}=0,
\end{equation}

\begin{equation}
P_{x}\left\{ Z_{t} \geq 0, \forall t \right \}=1, \forall x\geq 0,
\label{Z0}
\end{equation}

\begin{equation}
E_{x}\left[ \int_{0}^{\infty}e^{-\lambda_i t}dL \right] < \infty, \forall x\geq 0,\,\, i=1,2,
\label{Lintegrab}
\end{equation}

and
\begin{equation}
E_{x}\left[ \int_{0}^{\infty}e^{-\lambda_1 t}dU \right] < \infty, \forall x\geq 0.
\label{Uintegrab}
\end{equation}
We denote by ${\textit{S}}(x)$ the set of all feasible policies associated with the continuous process $X$ that starts at $x$.
\end{mydef}

 \subsection{Transaction Costs}

We assume that the bank can continuously sell and buy funds, thus lowering or increasing its excess reserve account.

It is considered, as in \cite{ChenMazumdar}, that there are three types of transaction costs:
\begin{enumerate}
 \item A proportional transaction cost $\alpha$ of buying funds. 
 
 \item A proportional transaction cost $\beta$ of selling federal funds.\footnote{The proportional adjustment costs, $\alpha$ and $\beta$, are due to `spreads between bid and ask prices, brokerage fees, the lack of availability of a transaction partner and other service charges which vary with the volume of the transaction', as in \cite{ChenMazumdar}and  \cite{Frost}.}
 
 \item A continuous holding cost, incurred at the rate $h.$
\end{enumerate}

 \section{The Problem Formulation}

 \subsection{The Cost Function}

 \begin{mydef}\label{d:k}
  The \textsl{cost function} associated to the feasible policy $(L,U)$ is 
  \begin{equation}
k_{L,U}(x)\equiv E_{x}\left[ \int_{0}^{\infty}[e^{-\lambda_1 t}(hZ_{t}dt+\beta dU) + (n e^{-\lambda_1 t} +(1-n) e^{-{\lambda_2} t})\alpha dL]        \right],\qquad  x\geq0,
\label{cost}
\end{equation}
with $n\in[0,1].$
\end{mydef}

\begin{rem}
 The cumulative funds purchases and funds sales are discounted at possible different rates. If $n=1$ then the discounting occur at the same rate $\lambda_1.$
 The discount function $n e^{-\lambda_1 t} +(1-n) e^{-{\lambda_2} t}, \, n\in[0,1]$ was considered in \cite{EkePir} and leads to a time changing discount rate
 in $[\lambda_2, \lambda_1].$
  
\end{rem}

 \subsection{The Objective}

    The bank's reserve management and profit-making problem is to find the optimal strategy $(\hat{L},\hat{U})$ which minimizes the cost. 

\begin{mydef}
 The control $(\hat{L},\hat{U})$ is said to be \textsl{optimal} if $k_{\hat{L},\hat{U}}(x)$ is minimal among the cost functions $k_{L,U}(x)$ associated with feasible policies $(L,U)$, for each fixed $x\geq 0$.
\end{mydef}
 
 The problem of minimizing the cost can be translated to the task of maximizing a value function. This function is easier to work with and it turns out to have particular characteristics, when the policy is of a barrier type. We present the relation between the cost function and the gain function obtained in \cite{Harrison}.

 \subsection{The Gain Function}
 \begin{mydef}
 The gain function is defined by
  \begin {equation}
  v_{L,U}(x) \equiv E_{x}\left\{\int_{0}^{\infty} e^{-\lambda_1 t}(rdU- cdL) \right\} - E_{x}\left\{\int_{0}^{\infty} e^{-\lambda_2 t}(1-n) \alpha dL  \right\},\qquad x\geq 0,
    \label{v}
  \end{equation}
  where $r \equiv h/\lambda_1 -\beta,$ and $c \equiv h/\lambda_1+ n \alpha$. 
  \end{mydef}
  Then extending the arguments from \cite{Harrison} one gets the following Lemma.
  
  \begin{lem}\label{11}
   The relation between the cost function and the gain function is
      \begin{equation}\label{kv}
  k_{L,U}(x)= hx/\lambda_1 + h\mu/\lambda_{1}^{2}- v_{L,U}(x), x\geq0.
  \end{equation}
 \end{lem}
 
 \section{The Optimal Policies}
 
 \subsection{The Barrier Policies}
 
    Let $b>0$ be a real fixed number. We consider that $X_{0}=x\in [0,b]$. If $X_{0}>b$, then we allow a jump at $0$ for $U$: $U_{0}=X_{0}-b$.\\

   \begin{mydef}\label{Sc}
     The barrier policies are the set of policies $(L,U)\in \textit{S}(x)$ that satisfy:
     \begin{enumerate}
      \item  $(L,U)$ continuous on $(0,\infty)$, increasing, $L_{0-}=U_{0-}=0$,
      
       \item $Z_{t}\equiv X_{t}+L_{t}-U_{t}\geq 0, \forall t\geq 0$,
        
      \item $ \int_{0}^{t}I_{Z_{t}>0}dL_{t}=0, \int_{0}^{t}I_{Z_{t}<b}dU_{t}=0.$ 
      \end{enumerate} 
   \end{mydef}
   
   A barrier policy $(L,U)$ satisfies:
\begin{equation}\label{LUsystem}
L_{t}=\sup_{0\leq s\leq t}(X_{s}-U_{s})^{-}, U_{t}=\sup_{0\leq s\leq t}(b-X_{s}-L_{s})^{-},
\end{equation}
where $x^{-}$ denotes the negative part of $x$. Moreover, the Double Skorokhod Formula obtained in \cite{Kavita}  can be translated into a formula for the bank's  transaction amount $L-U$, as shown in \cite{CanepaPirvu}:

  \begin{equation} \label{XLU}
L_{t} - U_{t} = -[(X_{0}-b)^{+}\wedge \inf_{u\in [0,t]}X_{u}] \vee \sup_{s\in [0,t]}[(X_{s}-b)\wedge \inf_{u\in[s,t]}
X_{u}].
\end{equation}
 
  \subsection{The Optimal Policy}
 
 Let $-\gamma_{1}, \bar{\gamma}_{1}$ be the roots of $ \sigma^{2}\gamma^{2}/2 + \mu \gamma - \lambda_1 =0,$
 \begin{equation}
\label{alpha_}
\gamma_{1}\equiv \frac{\sqrt{\mu^{2}+2\sigma^{2}\lambda_1} + \mu}{\sigma^{2}}>0,
\end{equation}
\begin{equation}
\label{alpha^}
\bar{\gamma}_{1}\equiv \frac{\sqrt{\mu^{2}+2\sigma^{2}\lambda_1} - \mu}{\sigma^{2}}>0. 
\end{equation}
Define 
\begin{equation}\label{g}
g(x)\equiv \gamma_{1}e^{\bar{\gamma}_{1} x} + \bar{\gamma}_{1}e^{-\gamma_{1}x}. 
\end{equation}
Then $g(0)>0,$ $g'(0)=0,$ and $g$ is strictly decreasing and continuous on $(-\infty, 0]$. Hence there must be a point $-b<0$ such that 
\begin{equation}\label{b}
g(-b)=g(0)c/r.
\end{equation}

Let $\gamma_{2}$ the positive root of $ \sigma^{2}\gamma^{2}/2 + \mu \gamma - \lambda_2 =0,$
\begin{equation}
\label{alpha^_}
\gamma_{2}\equiv \frac{\sqrt{\mu^{2}+2\sigma^{2}\lambda_2} - \mu}{\sigma^{2}}>0. 
\end{equation}
Define 
\begin{equation}
  \label{v0b}
  v_{1}(x) \equiv
\left\{
	\begin{array}{ll}
		 \frac{r}{g'(b)}g(x)+ \frac{c }{g'(-b)}g(x-b) & \mbox{if } 0\leq x\leq b \\
		 v_{1}(b)+ (x-b)r & \mbox{if }  x>b.
	\end{array}\right.
\end{equation}
 
 and
 
 \begin{equation}\label{v2}
 v_{2}(x) \equiv   -\frac{(1-n) \alpha   }{  \gamma_{2}  } e^{-\gamma_{2}x}. 
\end{equation}
 
  \begin{prop}
  \label{BarrierValue}
  The barrier policy $(\hat{L}, \hat{U})$ associated with $b$ of \eqref{b} is admissible, i.e., $(\hat{L}, \hat{U})\in \textit{S}(x).$ Moreover
   \begin {equation}
  v_{1}(x) = E_{x}\left\{\int_{0}^{\infty} e^{-\lambda_1 t}(rd\hat{U}- cd \hat{L}) \right\},\qquad x\geq 0,
    \label{v1}
  \end{equation}
   \begin {equation}
  v_{2}(x) = - E_{x}\left\{\int_{0}^{\infty} e^{-\lambda_2 t}(1-n) \alpha d \hat{L}  \right\},\qquad x\geq 0.
    \label{v2}
    \end{equation}
   Therefore
   \begin {equation}
  v_{\hat{L},\hat{U}}(x)= v_{1}(x) + v_{2}(x).
    \label{v3}
  \end{equation}
  
  \end{prop}

  \subsection{Main Result}
 
 The following is the main result of our paper.
 \begin{thm}\label{main}
  The barrier policy $(\hat{L}, \hat{U})$ associated with $b$ of \eqref{b} is optimal, i.e., for every $({L}, {U})\in \textit{S}(x),$
   \begin {equation}
  v_{L,U}(x)\leq v_{  \hat{L}, \hat{U}}(x).
    \label{v4}
   \end{equation}
 \end{thm}

%Given an upper barrier $b$, the algorithm \ref{algLU0} provides the controlled process $Z=X+L-U$.
% \begin{algorithm}
% \caption{Algorithm for computing the controlled process, given the path of the underlying and the upper barrier}
% \label{algLU0}
%\begin{algorithmic}[1]
%\STATE{\bf{Input:}$(X_{s})_{s=0,N},b$
%\STATE{\bf{Output: $(Z_{s})_{s=0,N},(L_{s})_{s=0,N},(U_{s})_{s=0,N}$} }
%\STATE $L_{0},U_{0} \leftarrow 0$
%\IF {$X_{0}\leq 0$}
%\FOR {$s=0$ to $N$} 
%\STATE $X_{s} \leftarrow X_{s}-X_{0}+0.1$
%\ENDFOR
%%\ENDIF
%%
%\IF {$X_{0}\geq b$}
%%\THEN
%\STATE $U_{0} \leftarrow X_{0}-b$
%\FOR {$s=0$ to $N$} 
%\STATE $X_{s} \leftarrow X_{s} -U_{0}$
%\ENDFOR
%%\ENDIF
%\STATE $Z_{0} \leftarrow X_{0}$ 
%\IF {$0<X_{0}\leq b$}
%%\THEN
%\FOR {$t=0$ to $N$}
%\IF {$X_{t+1}-X_{t}+Z_{t} \geq 0$}
%%\THEN
%\STATE $L_{t+1}=L_{t}$
%\ELSE $L_{t+1}= \max(L_{t}, -(X_{t+1}-U_{t}))$
%%\ENDIF
%\IF {$X_{t+1}-X_{t}+Z_{t} \leq b$}
%\STATE $U_{t+1}=U_{t}$
%\ELSE $U_{t+1}= \max(U_{t}, X_{t+1}+L_{t}-b)$
%%\ENDIF
%\STATE $Z_{t+1} \leftarrow X_{t+1}+L_{t+1}-U_{t+1}$
%\ENDFOR  
%%\ENDIF
%\end{algorithmic}
%\end{algorithm} 
%% 

 \subsection{Special Case}
 
 Let us take $n=1$ so that we have same discount rate $\lambda_1.$ Moreover,
 let the drift $\mu$ and volatility $\sigma$ depend on the bank size $A.$ Inspired by  \cite{CanepaPirvu}
 we take $\mu$ and $\sigma$ linear in $A,$ i.e., $\mu=k_1 A, \,\, \sigma= k_2 A.$ 
 
 \begin{corollary}\label{lin}
 The barrier $b$ is linear in the bank size $A.$ Consequently, $L-U$ is increasing in the bank size $A.$
 \end{corollary}
 
 \begin{rem}
 This result can be used by a bank to develop a strategy for selling funds when its controlled excess reserve process hits this upper optimal barrier b, i.e. a certain percent of its asset's size. This corollary is consistent with the so called small bank large bank dichotomy, meaning that the bigger the size of the bank, the larger the net purchase amount that the bank undertakes.
 \end{rem}

 \section{Appendix}
 
   \subsection{Appendix A: Proof of Lemma \ref{11}} \begin{proof} Wlog, we can assume that $U_{0}=L_{0}=0$ (the other cases are similar, given (\ref{Lintegrab})). Since $Z\equiv X + L - U$,  we have:
\begin{equation}
h E_{x} \left(\int_{0}^{\infty}e^{-\lambda_1 t}Z_{t}dt \right)= hE_{x} \left(\int_{0}^{\infty}e^{-\lambda_1 t}X_{t}dt \right)+ hE_{x} \left(\int_{0}^{\infty}e^{-\lambda_1 t}(L_{t}-U_{t})dt \right).
\label{kv1}
\end{equation}
Applying Fubini's theorem, we obtain:
$$E_{x} \left(\int_{0}^{\infty}e^{-\lambda_1 t}X_{t}dt \right) = \int_{0}^{\infty}e^{-\lambda_1 t}E_{x}(X_{t})dt.$$
We know that since $X$ is a $(\mu, \sigma)$ Brownian motion, $E_{x}(X_{t}) = x+\mu t$ and a simple integration gives the following:
$$\int_{0}^{\infty}e^{-\lambda_1 t}E_{x}(X_{t})dt = x/\lambda_1 + \mu/\lambda_1^{2}.$$
From the last two formulas we conclude that
\begin{equation}
E_{x} \left(\int_{0}^{\infty}e^{-\lambda_1 t}X_{t}dt \right)= x/\lambda_1 + \mu/\lambda_1^{2}.
\label{ElambdaX}
\end{equation}
Next, we recall the Riemann- Stieltjes integration by parts theorem, which states that if two functions $f,g$ are $FV$ (of finite variation), then:
$$\int_{0}^{t}f dg = f(t)g(t) - f(0)g(0)- \int_{0}^{t}g(s)df(s).$$ 
Noticing that since $L$ is increasing,  $L$ is $FV$ and applying the above-mentioned theorem, we obtain, for each fixed $T>0$ and for general $\lambda>0$:
\begin{equation}
\int_{0}^{T}e^{-\lambda t }dL = e^{-\lambda T}L_{T}+ \lambda \int_{0}^{T}e^{-\lambda t}L_{t}dt.
\label{LintParts}
\end{equation}

Applying Fatou's lemma twice, (\ref{LintParts}) and (\ref{Lintegrab}), we obtain
\begin{align*}
E_{x}[\liminf_{T \rightarrow \infty}(e^{-\lambda T}L_{T}+ \lambda \int_{0}^{T}e^{-\lambda t}L_{t}dt)]&=E_{x}[\liminf_{T \rightarrow \infty}E_{x}\int_{0}^{T}e^{-\lambda t }dL] \leq \liminf_{T \rightarrow \infty}E_{x}\int_{0}^{T}e^{-\lambda t }dL\\\nonumber
& \leq \limsup_{T \rightarrow \infty}E_{x}\int_{0}^{T}e^{-\lambda t }dL \leq E_{x}\limsup_{T \rightarrow \infty} \int_{0}^{T}e^{-\lambda t }dL\\\nonumber
& = E_{x} \int_{0}^{\infty}e^{-\lambda t}dL <\infty .\nonumber
\end{align*}

 It follows that $e^{-\lambda t}L_{t}\rightarrow 0$ almost surely as $t\rightarrow \infty$. Indeed, if this were not true, then, since $e^{-\lambda t}L_{t} \geq 0$ on a set of non-zero measure, we would have $\int_{0}^{\infty}e^{-\lambda t}L_{t}dt=\infty$. We obtain therefore that\\
  $E_{x}[\liminf_{T \rightarrow \infty}(e^{-\lambda T}L_{T}+ \lambda \int_{0}^{T}e^{-\lambda t}L_{t}dt)]$ becomes unbounded, and we get to a contradiction.
  
 Letting $T\rightarrow \infty$ in (\ref{LintParts})  and then taking $E_{x}$ on both sides, we obtain for every $\lambda>0$:
 \begin{equation}
 E_{x} \left(\int_{0}^{\infty}e^{-\lambda t}L_{t}dt \right)= \frac{1}{\lambda}E_{x}\left( \int_{0}^{\infty}e^{-\lambda t }dL \right).
  \label{Lform}
 \end{equation} 
 We obtain a similar equation for $U$. Replacing it and (\ref{Lform}) (for both $\lambda_1$ and $\lambda_2$), (\ref{ElambdaX}), (\ref{kv1}) in the definition (\ref{cost}) for $k_{L,U}(x)$, we obtain:
 \begin{eqnarray*}
 k_{L,U}(x) &=& h(\frac{x}{\lambda_1}+\frac{\mu}{\lambda_1^{2}})+ (\frac{h}{\lambda_1}+ n\alpha)E_{x}\left(\int_{0}^{\infty}e^{-\lambda_1 t}dL_{t}\right) +(-\frac{h}{\lambda_1}+ \beta)E_{x}\left(\int_{0}^{\infty}e^{-\lambda_1 t}dU_{t}\right)\\&+&(1-n)E_{x}\left(\int_{0}^{\infty}e^{-\lambda_2 t}\alpha dL_{t}\right).
 \end{eqnarray*}
\end{proof}

  \subsection{Appendix B: Proof of Proposition \ref{BarrierValue}}

  \begin{proof}
  
If $u:R\rightarrow R$ is a function of class $C^{2}$ (i.e. twice continuously differentiable), we denote by $\Gamma$ the generator of the continuous diffusion process $X$ in (\ref{X}):
\begin{equation}\label{Gamma}
\Gamma u = \mu u'+\frac{\sigma^{2}}{2}u''.
\end{equation}
 Since $Z=X+\LL-\UU$ then
\begin{equation}\label{R2}
 d v_{1}(Z_{t})=\sigma v_{1}'(Z)dB_{t} + [\Gamma v_{1}(Z)dt + v_{1}'(0)d\LL - v_{1}'(b)d\UU].
 \end{equation}
 Indeed by Ito's Lemma combined with the fact that $\LL$ increases only when $Z=0$, whereas $\UU$ increases only when $Z=b$ yields:
 \begin{align*}
 dv_{1}(Z_{t})&= v_{1}'(Z_{t})dZ_{t}+ 1/2v_{1}''(Z_{t})(dZ_{t})^{2}\\
  &= v_{1}'(Z)(dX + d\LL- d\UU)+ \frac{1}{2}\sigma^{2}v_{1}''(Z)dt\\
  &=v_{1}'(Z)(\mu dt+\sigma dB_{t} + d\LL_{t}-d\UU_{t})+\frac{1}{2}\sigma^{2}v_{1}''(Z)dt\\
  &= \sigma v_{1}'(Z)dB_{t}+ \Gamma v_{1}(Z)dt + v_{1}'(0)d\LL-v_{1}'(b)d\UU.
 \end{align*}
 Since $Z$ is bounded a.s. then $e^{-\lambda s}v_{1}'(Z)$ is bounded a.s., thus the process
 $M_{t}\equiv \int_{0}^{t}e^{-\lambda s}v_{1}'(Z_{t})\sigma dB_{t}, t\geq 0$ is a martingale. Consequently
 $$E_{x}M_{t}=0.$$
 From the definition of $ v_{1}$ we infer that
 $$\Gamma v_1(z) = \lambda_1 v_1(z),\, z\in[0,b],\quad  v_{1}'(0)=c,\quad   v_{1}'(b)=r.   $$
  Applying integration by parts leads to:
 \begin{align*}
 e^{-\lambda_1 t}v_{1}(Z_{t}) &= v_{1}(Z_{0}) + \int_{0}^{t}e^{-\lambda_1 s}dv_{1}(Z) - \lambda_1 \int_{0}^{t}e^{- \lambda_1 s}v_{1}(Z)ds\\
 &=v_{1}(Z_{0}) + M_{t} +\int_{0}^{t}e^{- \lambda_1 s}[\Gamma v_{1}(Z)ds+ v_{1}'(0)d\LL-v_{1}'(b)d\UU]- \lambda_1\int_{0}^{t}e^{- \lambda_1 s} v_{1}(Z)ds \\
 &=v_{1}(Z_{0}) + M_{t} +\int_{0}^{t}e^{- \lambda_1 s}[\Gamma v_{1}(Z)-\lambda_1 v_{1}(Z)]ds - \int_{0}^{t}e^{- \lambda_1 s}[v_{1}'(b)d\UU- v_{1}'(0)d\LL]\\
  &=v_{1}(Z_{0})+ M_{t}- \int_{0}^{t}e^{- \lambda_1 s}[r d\UU-cd\LL].
  \end{align*}
  
  By taking expectation, then letting $t\rightarrow\infty,$ and using that $v_{1}$ is bounded leads to
  
  \begin{equation}\label{e0}
  v_{1}(x) = E_{x}\left\{\int_{0}^{\infty} e^{-\lambda_1 t}(rd\hat{U}- cd \hat{L}) \right\}.
  \end{equation}
  
  Let $\ZZ=X+\LL$ then
\begin{equation}\label{R2}
 d v_{2}(\ZZ_{t})=\sigma v_{2}'(\ZZ)dB_{t} + [\Gamma v_{2}(\ZZ)dt + v_{2}'(0)d\LL].
 \end{equation}
 Indeed by Ito's Lemma combined with the fact that $\LL$ increases only when $Z=0$ yields:
 \begin{align*}
 dv_{2}(\ZZ_{t})&= v_{2}'(\ZZ_{t})d\ZZ_{t}+ 1/2v_{2}''(\ZZ_{t})(d\ZZ_{t})^{2}\\
  &= v_{2}'(\ZZ)(dX + d\LL)+ \frac{1}{2}\sigma^{2}v_{2}''(\ZZ)dt\\
  &=v_{2}'(\ZZ)(\mu dt+\sigma dB_{t} + d\LL_{t})+\frac{1}{2}\sigma^{2}v_{2}''(\ZZ)dt\\
  &= \sigma v_{2}'(\ZZ)dB_{t}+ \Gamma v_{2}(\ZZ)dt + v_{2}'(0)d\LL.
 \end{align*}
 Since $\ZZ$ is positive a.s. then $e^{-\lambda s}v_{2}'(\ZZ)$ is bounded a.s., thus the process
 $N_{t}\equiv \int_{0}^{t}e^{-\lambda s}v_{2}'(\ZZ_{t})\sigma dB_{t}, t\geq 0$ is a martingale. Consequently
 $$E_{x}N_{t}=0.$$
 From the definition of $ v_{2}$ we infer that
 $$\Gamma v_2 = \lambda_2 v_2,\quad  v_{2}'(0)= (1-n)\alpha.$$
  Applying integration by parts leads to:
 \begin{align*}
 e^{-\lambda_2 t} v_{2}(\ZZ_{t}) &= v_{2}(\ZZ_{0}) + \int_{0}^{t}e^{-\lambda_2 s}dv_{2}(\ZZ) - \lambda_2 \int_{0}^{t}e^{- \lambda_2 s}v_{2}(\ZZ)ds\\
 &=v_{2}(\ZZ_{0}) + N_{t} +\int_{0}^{t}e^{- \lambda_2 s}[\Gamma v_{2}(\ZZ)ds+ v_{2}'(0)d\LL]- \lambda_2\int_{0}^{t}e^{- \lambda_2 s} v_{2}(\ZZ)ds \\
 &=v_{2}(\ZZ_{0}) + N_{t} +\int_{0}^{t}e^{- \lambda_2 s}[\Gamma v_{2}(\ZZ)-\lambda_2 v_{2}(\ZZ)]ds - \int_{0}^{t}e^{- \lambda_2 s}[- v_{2}'(0)d\LL]\\
  &=v_{2}(Z_{0})+ N_{t}- \int_{0}^{t}e^{- \lambda_2 s}[(n-1)\alpha d\LL].
  \end{align*}
  
  By taking expectation, then letting $t\rightarrow\infty,$ and using that $v_{1}$ is bounded on $R^+$ leads to
  
  \begin{equation}\label{e1}
   v_{2}(x) = - E_{x}\left\{\int_{0}^{\infty} e^{-\lambda_2 t}(1-n) \alpha d \hat{L}  \right\}.
\end{equation}

Next let us check the feasibility of the policy $(\LL, \UU),$ i.e.,

\begin{equation}
E_{x}\left[ \int_{0}^{\infty}e^{-\lambda_i t}d\LL \right] < \infty, \forall x\geq 0,\,\, i=1,2,
\label{Lintegrab1}
\end{equation}

and
\begin{equation}
E_{x}\left[ \int_{0}^{\infty}e^{-\lambda_1 t}d\UU \right] < \infty, \forall x\geq 0.
\label{Uintegrab1}
\end{equation}

From \eqref{e1} we get 

\begin{equation}
E_{x}\left[ \int_{0}^{\infty}e^{-\lambda_2 t}d\LL \right] < \infty.
\label{Lintegrab1}
\end{equation}

Moreover

\begin{equation}
E_{x}\left[ \int_{0}^{\infty}e^{-\lambda_1 t}d[\LL-\UU] \right] = E_{x}\left[ \int_{0}^{\infty}e^{-\lambda_1 t}d[Z-X] \right] < \infty, \forall x\geq 0,
\label{Uintegrab11}
\end{equation}
combined with \eqref{e0} yield

\begin{equation}
E_{x}\left[ \int_{0}^{\infty}e^{-\lambda_1 t}d\LL \right] < \infty,\qquad E_{x}\left[ \int_{0}^{\infty}e^{-\lambda_1 t}d\UU \right] < \infty. 
\label{Lintegrab1}
\end{equation}
\end{proof}

 \subsection{Appendix C: Proof of Theorem \ref{main}}
 \begin{proof}
The idea of the proof is based on the martingale/supermartingale principle.
In a first step we show that some processes are supermartingales.
\begin{lem}\label{supermg}
For every $ (L,U) \in{\textit{S}}(x),$
with $Z=X+L-U,$ the process
$$e^{-\lambda_1 t}v_{1}(Z_{t}) +  \int_{0}^{t}e^{- \lambda_1 s}[r d U-cdL], \quad t\geq 0$$
is supermartingale. Moreover with $\ZZ=X+L,$ the process 
$$ e^{-\lambda_2 t}v_{2}(\ZZ_{t}) + \int_{0}^{t}e^{- \lambda_2 s}[(n-1)]\alpha dL,, \quad t\geq 0 $$  
 is supermartingale.
\end{lem}

 Therefore for a fixed $T>0,$ by taking expectations we get
 $$ E_{x}\left\{\int_{0}^{T} e^{-\lambda_1 t}(rd {U}- cd {L}) \right\} \leq  v_{1}(x) -  E_{x}[  e^{-\lambda_1 T} v_{1}(Z_{T})].$$ 
 Next, the positivity of $Z,$ the linearity of $v_{1}(z)$ for large $z,$ the integrability conditions \eqref{Lintegrab}, \eqref{Uintegrab}
  and Dominated Convergence Theorem yield that
   $$ E_{x}\left\{\int_{0}^{\infty} e^{-\lambda_1 t}(rd {U}- cd {L}) \right\} \leq  v_{1}(x).$$ 
   Similarly 
    $$ E_{x}\left\{\int_{0}^{T} e^{-\lambda_2 t}(n-1)d {L}) \right\} \leq  v_{2}(x) -  E_{x}[  e^{-\lambda_2 T} v_{2}(\ZZ_{T})].$$ 
    The positivity of $\ZZ,$ the roundedness of $v_{1}(z)$ for positive $z,$ the integrability condition \eqref{Lintegrab}.
  and Dominated Convergence Theorem yield that
   $$ E_{x}\left\{\int_{0}^{\infty} e^{-\lambda_2 t}(n-1)d {L}) \right\} \leq  v_{2}(x) .$$ By adding these inequalities we get
   $$  E_{x}\left\{\int_{0}^{\infty} e^{-\lambda_1 t}(rd {U}- cd {L}) \right\}-E_{x}\left\{\int_{0}^{T} e^{-\lambda_2 t}(1-n)d {L}) \right\} \leq  v_{1}(x) + v_{2}(x).$$
   However by Proposition \ref{BarrierValue}
  $$ v_{\hat{L},\hat{U}}(x)= v_{1}(x) + v_{2}(x),$$
   which proves optimality of $(\hat{L},\hat{U})$
    \subsubsection{Proof of Lemma \ref{supermg}}
   Recall that
   \begin{equation}\label{tt}
   \Gamma v_1(z) = \lambda_1 v_1(z),\, z\in[0,b],\quad  v_{1}'(0)=c,\quad   v_{1}'(b)=r.
   \end{equation}
   Moreover
   \begin{equation}\label{ttt}
    \Gamma v_1(z) \leq \lambda_1 v_1(z),\quad  r \leq v'_1(z)\leq c,\qquad z\geq 0.
   \end{equation}
   By Ito's Lemma for processes with jumps
   \begin{align*}
   d\left(e^{-\lambda_1 t}v_{1}(Z_{t}) +  \int_{0}^{t}e^{- \lambda_1 s}[r d U-cdL] \right)   &=
 e^{-\lambda_1 t}(\Gamma v_1 -\lambda v_1)(Z) dt\\
  &+  e^{-\lambda_1 t}(v_{1}'(Z)-c) d\tilde{L}+ e^{-\lambda_1 t}(r-v_{1}'(Z)) d\tilde{U}\\
  &+ \sum_{0\leq s\leq t}e^{-\lambda s}(\Delta v_1(Z)_{s}-c\Delta L_{s}+r \Delta U_{s}),
 \end{align*}
   where $d\tilde{L}=dL-\Delta L$ and $d\tilde{U}=dU-\Delta U.$
   In the light of \eqref{ttt} the claim yield if we prove that 
   $$  \sum_{0\leq s\leq t}e^{-\lambda s}(\Delta v_1(Z)_{s}-c\Delta L_{s}+r \Delta U_{s})\leq 0. $$
   Suppose that $\Delta L_{t}>0$ and $\Delta U_{t}=0 $ (the other cases are similar). Then $\Delta Z_{t}= \Delta L_{t}$ and
 $$\Delta v_1 (Z)_{t}-c\Delta L_{t}+r \Delta U_{t}=v_{1} (Z_{t})- v_{1} (Z_{t}-\Delta L_{t})-c \Delta L_{t}.$$
 The last quantity is negative because $v_1'(z)\leq c,\,z\geq 0.$
 Recall that
 \begin{equation}\label{tttt}
 \Gamma v_2 = \lambda_2 v_2,\quad  v_{2}'(0)= (1-n)\alpha.
 \end{equation}
   
   Moreover
   \begin{equation}\label{ttttt}
  v_{2}'(z)\leq (1-n)\alpha,\qquad z\geq 0.
   \end{equation}
    By Ito's Lemma for processes with jumps
   \begin{align*}
   d\left(e^{-\lambda_2 t}v_{2}(\ZZ_{t}) + \int_{0}^{t}e^{- \lambda_2 s}[(n-1)]\alpha dL\right)
    &= e^{-\lambda_2 t}(\Gamma v_2 -\lambda v_1)(\ZZ) dt\\
  &+  e^{-\lambda_2 t}(v_{2}'(\ZZ)-(1-n)\alpha) d\tilde{L}+\\
  &+ \sum_{0\leq s\leq t}e^{-\lambda s}(\Delta v_2(\ZZ)_{s}-(1-n)\alpha\Delta L_{s}),
 \end{align*}
In the light of \eqref{ttttt} the claim yield if we prove that 
   $$  \sum_{0\leq s\leq t}e^{-\lambda s}(\Delta v_2(\ZZ)_{s}-(1-n)\alpha\Delta L_{s})\leq 0. $$
   Suppose that $\Delta L_{t}>0,$ then $\Delta \ZZ_{t}= \Delta L_{t}$ and
 $$\Delta v_2 (Z)_{t}-(1-n)\alpha\Delta L_{t}=v_{2} (Z_{t})- v_{2} (Z_{t}-\Delta L_{t})-(1-n)\alpha \Delta L_{t}.$$
 The last quantity is negative because $v_2'(z)\leq (1-n)\alpha,\,z\geq 0.$
\end{proof}

  \subsection{Appendix D: Proof of Corollary \ref{lin}}
  
   \begin{proof}
    In light of $\mu=k_1 A, \,\, \sigma= k_2 A, $ and 
     \begin{equation}
\label{alpha_1}
\gamma_{1}\equiv \frac{\sqrt{\mu^{2}+2\sigma^{2}\lambda_1} + \mu}{\sigma^{2}},
\end{equation}
\begin{equation}
\label{alpha^1}
\bar{\gamma}_{1}\equiv \frac{\sqrt{\mu^{2}+2\sigma^{2}\lambda_1} - \mu}{\sigma^{2}}, 
\end{equation}
it follows that $\gamma_{1}=a_1 A, \,\, \bar{\gamma}_{1}= \bar{a}_1 A,$ for some constants ${a}_1, \bar{a}_1.$ 
Recall that
$$
g(x)\equiv \gamma_{1}e^{\bar{\gamma}_{1} x} + \bar{\gamma}_{1}e^{-\gamma_{1}x},
$$
whence
$$ \frac{g(-b)}{g(0)} =F\left(\frac{b}{A}\right),$$
for some function $F.$ From \eqref{b} it follows that $b$ should solve
\begin{equation}\label{bb}
F\left(\frac{b}{A}\right)=c/r.
\end{equation}
Therefore $b=k A,$ for some positive constant $k.$

Given $X_{0}\in[0,b]$, then $(X_{0}-b)^{+}=0$, The Double Skorokhod Formula (see \cite{Kavita}) yields
that $L-U$ is increasing in the barrier $b.$ Since $b=k A,$ for some positive constant $k,$  
then $L-U$ is increasing in the bank size $A.$
\end{proof}

\end{document}